\documentclass[10pt,final,journal,twocolumn]{IEEEtran}

\ifCLASSINFOpdf

\else

\fi

\usepackage{cite}
\usepackage[cmex10]{amsmath}
\usepackage{amsmath}
\usepackage{algorithmic}
\usepackage{stfloats}
\usepackage{multicol,multienum}
\usepackage{multirow}
\usepackage[dvips]{graphicx}

\usepackage[tight,footnotesize]{subfigure}
\usepackage{wrapfig}
\usepackage{color}

 \usepackage{booktabs}
 \usepackage{multirow}

\usepackage{mathtools}
\usepackage{color}
\usepackage{float}
\usepackage{bm}

\usepackage{pifont}

\usepackage[ruled,linesnumbered]{algorithm2e}

\usepackage{bigfoot}

\begin{document}

\title{Generative AI for Secure and Privacy-Preserving Mobile Crowdsensing}

\author{Yaoqi~Yang,
Bangning~Zhang, Daoxing~Guo,
Hongyang~Du,
Zehui~Xiong,~\IEEEmembership{Senior Member,~IEEE},
Dusit~Niyato,~\IEEEmembership{Fellow,~IEEE},
and Zhu~Han,~\IEEEmembership{Fellow,~IEEE}

\thanks{Manuscript received xxx. }

\thanks{Yaoqi~Yang, Bangning~Zhang, and Daoxing~Guo are with the College of Communications Engineering, Army Engineering University of PLA, Nanjing 210000, China. (e-mail: yaoqi$\_$yang@yeah.net; bangning$\_$zhang@sina.com; xyzgfg@sina.com)}

\thanks{Hongyang Du and Dusit Niyato are with the School of Computer Science and Engineering, NTU, Singapore. (e-mail: hongyang001@e.ntu.edu.sg; dniyato@ntu.edu.sg)}

\thanks{Zehui Xiong is with the Information Systems Technology and Design Pillar, Singapore University of Technology and Design, Singapore. (e-mail: zehui$\_$xiong@sutd.edu.sg)}

\thanks{Zhu Han is with the Department of Electrical and Computer Engineering at the University of Houston, Houston, TX 77004, USA. (e-mail: hanzhu22@gmail.com)}

}

\maketitle

\begin{abstract}
Recently, generative AI has attracted \textcolor[rgb]{0.00,0.00,0.00}{much} attention from both academic and industrial fields, which has shown its potential, especially in the data generation and synthesis aspects. Simultaneously, secure and privacy-preserving mobile crowdsensing (SPPMCS) has been widely applied in data collection/acquirement due to an advantage on low deployment cost, flexible implementation, and \textcolor[rgb]{0.00,0.00,0.00}{high adaptability}. \textcolor[rgb]{0.00,0.00,0.00}{Since generative AI can generate new synthetic data to replace the original data to be analyzed and processed, it can lower data attacks and privacy leakage risks for the original data. Therefore, integrating generative AI into SPPMCS is feasible and significant.}
\textcolor[rgb]{0.00,0.00,0.00}{Moreover, this} paper investigates \textcolor[rgb]{0.00,0.00,0.00}{an} integration of generative AI in SPPMCS, where \textcolor[rgb]{0.00,0.00,0.00}{we present} potential research focuses, solutions, and case studies. Specifically, we firstly review the preliminaries for generative AI and SPPMCS, where their integration potential is presented. Then, we discuss research issues and solutions for generative AI-enabled SPPMCS, including security defense of malicious data injection, illegal authorization, malicious spectrum manipulation at the physical layer, and privacy protection on sensing data content, sensing terminals' identification and \textcolor[rgb]{0.00,0.00,0.00}{location}. Next, \textcolor[rgb]{0.00,0.00,0.00}{we propose a} framework for sensing data content protection with generative AI, and simulations results have \textcolor[rgb]{0.00,0.00,0.00}{clearly demonstrated} the effectiveness of the \textcolor[rgb]{0.00,0.00,0.00}{proposed framework}. Finally, we present major research directions for generative AI-enabled SPPMCS.
\end{abstract}


\IEEEpeerreviewmaketitle

\section{Introduction}

With the development of wireless communication systems, emerging technologies, \textcolor[rgb]{0.00,0.00,0.00}{such as Metaverse}, large language model, and digital twins, are mainly driven by the secure and privacy-preserving sensing data. Simultaneously, to meet the demand of generation, collection, and \textcolor[rgb]{0.00,0.00,0.00}{processing of} the massive amounts of sensing data in the secure manner, the \textcolor{black}{secure and privacy-preserving mobile crowdsensing (SPPMCS)} \cite{yangppfo} has been \textcolor[rgb]{0.00,0.00,0.00}{introduced}.
Generally, SPPMCS is composed of \textcolor[rgb]{0.00,0.00,0.00}{a} service requester (SR), service provider (SP), and mobile sensing terminals (MSTs) \cite{yangppfo}. Specifically, \textcolor[rgb]{0.00,0.00,0.00}{the} SR is responsible for sensing task generation, and handling \textcolor[rgb]{0.00,0.00,0.00}{specific} requirements to \textcolor[rgb]{0.00,0.00,0.00}{a certain} SP. \textcolor[rgb]{0.00,0.00,0.00}{On the one hand, the SP} can allocate the sensing tasks to MSTs after receiving the requirements. On the other hand, the \textcolor[rgb]{0.00,0.00,0.00}{SP} can integrate and process the collected data from \textcolor[rgb]{0.00,0.00,0.00}{the MSTs}. \textcolor[rgb]{0.00,0.00,0.00}{Here, the MSTs} can collect the needed sensing data through cooperation \textcolor[rgb]{0.00,0.00,0.00}{ with SP}.
In general, thanks to the integrated multi-technologies \textcolor[rgb]{0.00,0.00,0.00}{for}
wireless sensing, crowdsourcing, security enhancement, and privacy preservation, SPPMCS can ensure security of sensing data, and has been widely applied in \textcolor[rgb]{0.00,0.00,0.00}{a variety of applications due to its capable features} of low deployment cost, flexible implementation, and \textcolor[rgb]{0.00,0.00,0.00}{high adaptivity to various environments and user demands}.

Simultaneously, artificial intelligence (AI) technologies also make a great progress in recent years, where generative AI \cite{zhang2023generative} has been paid more and more attention in both academia and industry fields. \textcolor[rgb]{0.00,0.00,0.00}{Generative AI}\footnote{\textcolor[rgb]{0.00,0.00,0.00}{Note that AI-generated content (AIGC) focuses more on generating new and original content, while generative AI concentrates more on simulating or developing a particular data or phenomenon. The content generated by AIGC tends to be more innovative and unique, while the content generated by generative AI focuses more on simulation and realism. To meet the request of SR, the characteristics of the generated data should be kept as close as possible to those of the original data. Hence, this paper mainly investigates the integration between generative AI and SPPMCS.}} refers to technologies that generate content such as text, images, sound, video, \textcolor[rgb]{0.00,0.00,0.00}{and code}, based on specific algorithms, models, and rules. \textcolor{black}{For example, in healthcare management applications, generative AI can synthesize clinical notes for care managers, synthesize medical and referral information, and generate care plans and summaries for members\footnote{https://www.mckinsey.com/industries/healthcare/our-insights/tackling-healthcares-biggest-burdens-with-generative-ai}. In the education field, generative AI can help diverse learners with different learning abilities, linguistic backgrounds or accessibility needs, accelerate exploration and suggest new ideas\footnote{https://teaching.cornell.edu/generative-artificial-intelligence$\#$Upside}.}
\textcolor{black}{Variational autoencoder (VAE), generative adversarial network (GAN), generative diffusion model (GDM), and Transformer-based model (TBM)} have become \textcolor[rgb]{0.00,0.00,0.00}{exemplary} generative AI models, and they all have been widely used \textcolor[rgb]{0.00,0.00,0.00}{including text, image, and video generation, audio and sensing signal enhancement, etc.} For example, Murf (speech tools for voiceovers), Jasper AI (AI-driven copywriting tools), and Siri (smart assistant) are used to simplify the daily life and enhance the user experiences\footnote{https://murf.ai/resources/ai-use-in-everyday-life/}. Note that typical discriminative AI concentrates on handling existing data, e.g., classification, prediction and regression. Compared with the typical discriminative AI, \textcolor[rgb]{0.00,0.00,0.00}{generative AI is distinctive in the following aspects:}
\begin{itemize}
  \item \textbf{Data generation capability}. 
      Generative AI, as it contains a lot variables, has strong ability to create realistic, diverse, and high-quality contents. For example, in the GPT-4 application\footnote{https://openai.com/research/dall-e} of text to image, with text prompts ``flower", different flower images are generated in different interactions.
  \item \textbf{\textcolor{black}{Multimodal data adaptability.}} \textcolor{black}{Generative AI is adaptive to learn from a variety of data sources, which makes it can be applied in more scenarios. For example, Gemini\footnote{https://deepmind.google/technologies/gemini} can be trained and applied to generate synthetic text, image, audio and video data at the same time.}
  \item \textbf{Significant system performance gain}. As \textcolor[rgb]{0.00,0.00,0.00}{an} effective optimization problem solving tool, generative AI has shown \textcolor[rgb]{0.00,0.00,0.00}{its capability in significantly improving} system performance gain for the strategy-making, trend-prediction, and model-fitting processes\footnote{https://www.gartner.com/en/topics/generative-ai}.
\end{itemize}

\textcolor[rgb]{0.00,0.00,0.00}{An integration} of generative AI and SPPMCS \textcolor[rgb]{0.00,0.00,0.00}{will} further enhance the system's security and privacy preservation performance.
Through providing realistic synthetic data, detecting abnormal data \textcolor[rgb]{0.00,0.00,0.00}{flows}, and recognizing potential malicious attacks, generative AI can significantly enhance the defense capacity of SPPMCS. For example, when malicious attackers \textcolor[rgb]{0.00,0.00,0.00}{target} to infer identification information of MSTs by analyzing the MSTs' collected sensing data, generative AI, e.g., GAN and GDM, can help generate the synthetic sensing data, which can maintain nearly the same property of the real-world data to meet the requirements of SR. In this regard, by replacing the real-world generated data with relevant synthetic data, generative AI can conceal the identification information of MSTs, thus protecting the privacy of \textcolor[rgb]{0.00,0.00,0.00}{the} MSTs. Moreover, on the security enhancement aspect, the generated synthetic sensing data can be used for abnormal data flow detection. By training detection models with the synthetic data-enriched dataset, performance of the detector can be greatly improved.
In general, despite integrating SPPMCS and generative AI has the promising perspectives, the following specific issues \textcolor[rgb]{0.00,0.00,0.00}{have to} be well addressed.
\begin{itemize}
  \item \textbf{Q1}. Why do we need generative AI for SPPMCS? Specifically, on the one hand, what can generative AI do while non-AI and discriminative AI technologies cannot? On the other hand, which aspect can generative AI do better than non-AI and discriminative AI technologies?
  \item \textbf{Q2}. What are generative AI models suitable for SPPMCS?
  \item \textbf{Q3}. How can we apply generative AI to SPPMCS \textcolor[rgb]{0.00,0.00,0.00}{effectively}?
\end{itemize}

\textcolor[rgb]{0.00,0.00,0.00}{In this paper}, we aim at answering the \textcolor[rgb]{0.00,0.00,0.00}{above and other relevant} questions to \textcolor[rgb]{0.00,0.00,0.00}{demonstrate} the necessity, signification, importance, and forefront of the generative AI-enabled SPPMCS.
\textcolor{black}{Different from existing literature, we systematically demonstrate how generative AI is implemented in different data interaction stages within MCS. In addition, we evaluate how generative AI can address concerns of non-AI and discriminative AI-based approaches. Specifically, the major contributions of this paper can be summarized as follows.}
\begin{enumerate}
  \item We introduce the preliminaries for SPPMCS and generative AI. The comprehensive review about the two technologies will clearly identify the potential benefits of their integration. In addition, we also explore the advantages of generative AI in solving problems that non-AI and traditional discriminative AI cannot solve or handle well.
  \item We demonstrate possible scenarios and corresponding solutions about how to apply generative AI to SPPMSC, including security defense of malicious data injection, illegal authorization, malicious spectrum manipulation at the physical layer, and privacy protection on sensing data content, sensing terminals' identification and location. \textcolor[rgb]{0.00,0.00,0.00}{We} also point possible challenges to \textcolor[rgb]{0.00,0.00,0.00}{integrate} generative AI and SPPMCS.
  \item To evaluate the performance of generative AI for SPPMCS, we propose \textcolor[rgb]{0.00,0.00,0.00}{a} novel generative AI-enabled sensing data collection \textcolor[rgb]{0.00,0.00,0.00}{framework, where the GDM and GAN models} are used to realize privacy preservation of sensing data content. \textcolor{black}{Moreover, the numerical results evaluate that GAN can perform better than GDM in training porcess, while the quality of images generated by GDM is better than GAN.}
\end{enumerate}

The rest of the paper is arranged as follows. Section II introduces the preliminaries of SPPMCS and generative AI. Then, Section III presents how to integrate generative AI to SPPMCS, including research focuses, solutions, and potential challenges. Next, in Section IV, the framework for sensing data content protection is proposed for generative AI-enabled SPPMCS. Finally, Section V discusses some open issues.

\section{Preliminaries \textcolor[rgb]{0.00,0.00,0.00}{of} SPPMCS and Generative AI}


\subsection{Preliminaries \textcolor[rgb]{0.00,0.00,0.00}{of} SPPMCS}

Based on the crowdsourcing \textcolor[rgb]{0.00,0.00,0.00}{approach}, SPPMCS \textcolor[rgb]{0.00,0.00,0.00}{represents} the sensing data collection architecture, to obtain sensing data in a secure and privacy-preserving manner. SPPMCS integrates wireless sensing, crowsourcing, wireless attack defense, and privacy \textcolor[rgb]{0.00,0.00,0.00}{preservation technologies together in a unified fashion.}
\textcolor[rgb]{0.00,0.00,0.00}{In SPPMCS, the security issues mainly include malicious data injection, illegal authorization, and malicious spectrum manipulation at the physical layer. Privacy risk primarily focuses on the sensing data content, and MST's identification and location information. The system may face additional security and privacy issues in different sensing data interaction processes}, and the detailed process is presented as follows \cite{yangppfo}.


1) \textbf{Task Generation and Allocation.} In this stage, the SR generates \textcolor[rgb]{0.00,0.00,0.00}{a} sensing data collection request, including data type, amount, and quality, and then distribute it to \textcolor[rgb]{0.00,0.00,0.00}{an SP}. \textcolor[rgb]{0.00,0.00,0.00}{After} the SP receives the sensing task request from \textcolor[rgb]{0.00,0.00,0.00}{the SR}, it \textcolor[rgb]{0.00,0.00,0.00}{recruits} MSTs and allocate the sensing task to them. \textcolor[rgb]{0.00,0.00,0.00}{To ensure that} the sensing task can be successfully handled \textcolor[rgb]{0.00,0.00,0.00}{between} SP and MSTs, several non-AI methods have been adopted. For example, to prevent malicious \textcolor[rgb]{0.00,0.00,0.00}{attackers, e.g., beam-stealing attacks and man-in-the-middle attacks,} intercepting the sensing task and then allocating false one to MSTs, broadcast encryption was adopted \cite{yangppfo}. \textcolor{black}{Specifically, MSTs firstly register to the SP to evaluate the identification legitimacy, and obtain the private key for the sensing task. On this basis, after receiving the encrypted sensing task from SP, the authenticated MSTs can decrypt the task and help collect sensing data.} However, the sensing tasks need to be dynamically \textcolor[rgb]{0.00,0.00,0.00}{adjusted. For example, sensing data is needed from time to time. Thus, the encryption-based approach may encounter delays and incur key generation cost in adapting to sensing task content changes, degrading accuracy and efficiency of completing the sensing tasks.}

2) \textbf{Sensing Data Collection and Submission.} In this stage, after successfully confirming the sensing task content, MSTs can collaborative to finish sensing data collection tasks and then submitting to SP. In this regard, how to protect the sensing data content, location and identification information of MSTs can be an important issue, and some researches have adopted the traditional discriminative AI schemes for realizing the secure and privacy-preserving sensing data collection and submission. For example, federated learning is used to ensure the local data content privacy of MSTs \cite{wu2022eye}. However, traditional federated learning performance (i.e., convergence speed and model accuracy) is subject to the performance of local individuals (i.e., data amount, data quality, and transmission capacity). Due to the heterogeneity of MSTs in terms of storage, computing and communication capabilities, it is difficult to realize the fast convergence of SP, which brings challenges to real-time and high-accuracy application of SP.

\begin{table*}[t]
\centering
\caption{\textcolor{black}{Summarization and comparison for the typical generative AI models in SPPMCS.} } \label{Summarization}
\resizebox{0.9\width}{!}{
\begin{tabular}{c}
\includegraphics[width=20cm]{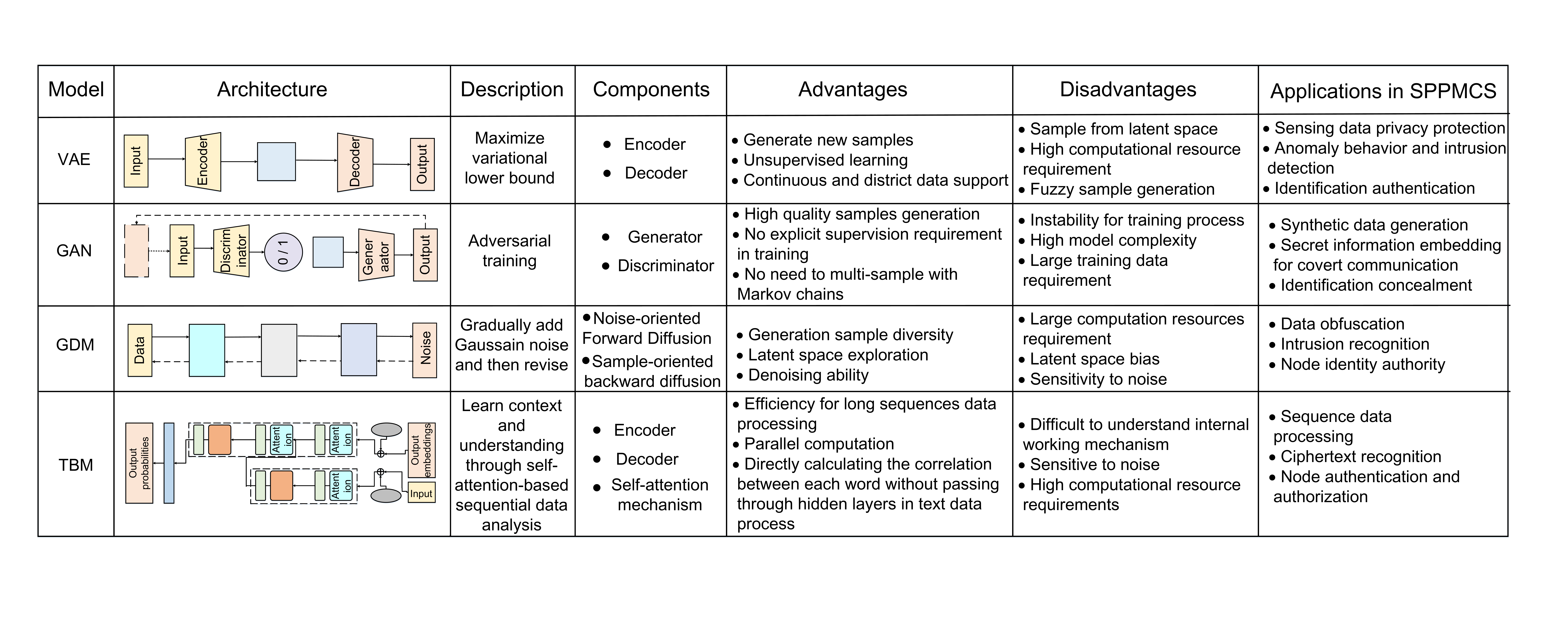}\\
\end{tabular}}
\end{table*}

3) \textbf{Result Evaluation and Reward Payment.} \textcolor[rgb]{0.00,0.00,0.00}{After the }SP receiving the collected sensing data from MSTs, the SP \textcolor[rgb]{0.00,0.00,0.00}{evaluates} quality of \textcolor[rgb]{0.00,0.00,0.00}{the} data to justify whether the data can meet the requirements of \textcolor[rgb]{0.00,0.00,0.00}{the SR} or not. \textcolor[rgb]{0.00,0.00,0.00}{After} checking the sensing data, \textcolor[rgb]{0.00,0.00,0.00}{the SP can} pay rewards to \textcolor[rgb]{0.00,0.00,0.00}{the} MSTs based on the evaluation results. In this regard, how to ensure the security of data evaluation process \textcolor[rgb]{0.00,0.00,0.00}{by an SP} is of great importance, and some non-AI and discriminative AI-based methods have been used. For example, \textcolor[rgb]{0.00,0.00,0.00}{a} reinforcement learning-based blockchain \cite{wang2020secure} approach \textcolor[rgb]{0.00,0.00,0.00}{was proposed}. However, due to the high computation cost and long convergence time of reinforcement learning, the transactions in blockchain cannot be verified and stored \textcolor[rgb]{0.00,0.00,0.00}{in real-time}.

Therefore, based on the above analysis and examples, it is significant to \textcolor[rgb]{0.00,0.00,0.00}{explore new AI technologies}, especially generative AI, to address the \textcolor[rgb]{0.00,0.00,0.00}{issues} that non-AI and traditional discriminative AI cannot well handle.

\subsection{Preliminaries \textcolor[rgb]{0.00,0.00,0.00}{of} generative AI}

Generative AI refers to a branch of AI systems that can perform functions similar to human creativity by learning from existing data and generating entirely new data. Generative AI can be implemented based on a variety of techniques, mainly including VAE, GAN, GDM, and TBM.

\textbf{Variational autoencoder (VAE).} Based on variational Bayesian inference, VAE, a generative model, can describe observations of the latent space from a probabilistic viewpoint. The VAE integrates the techniques of autoencoders and probabilistic coding to enable a more flexible and controllable process, where samples are generated by the modelling of latent representations. Unlike traditional autoencoders, VAE assumes that the latent representation obeys a prior distribution (usually a Gaussian distribution) and maps the input data to the distribution parameters of the latent space via an encoder. In SPPMCS, VAE can be applied in data augmentation. For example, in the sensing data submission process, to cope with the data imbalance \textcolor[rgb]{0.00,0.00,0.00}{issues} on intrusion detection (e.g., data injection attack) performance \cite{liu2022intrusion}, VAE \textcolor[rgb]{0.00,0.00,0.00}{was} adopted to realize data augmentation. \textcolor{black}{Moreover, the numerical results have shown that after enriching the training dataset with synthetic data, the F1-score of convolutional neural networks (CNN)-based intrusion detection model can improve 3.75\%.}

\textbf{Generative adversarial network (GAN).} GAN is a generative model that learns and evolves by two neural networks (i.e., \textcolor[rgb]{0.00,0.00,0.00}{a} generator and a discriminator) gaming each other. The generator takes random samples from the potential space as input, and its output needs to mimic the real samples in the training set as much as possible. After inputting the real samples or the output of the generator to \textcolor[rgb]{0.00,0.00,0.00}{the} discriminator, \textcolor[rgb]{0.00,0.00,0.00}{the discriminator} needs to separate the output of the generator from the real samples as much as possible. As a result, the generator and the discriminator \textcolor[rgb]{0.00,0.00,0.00}{compete} against each other and learn continuously. \textcolor[rgb]{0.00,0.00,0.00}{The} ultimate goal \textcolor[rgb]{0.00,0.00,0.00}{is to} enable the discriminator to judge whether the generator's output is true or not. In SPPMCS, GAN can be used to recognize the Sybil attacks. For example, in vehicular crowdsensing scenario \cite{jaimes2022generative}, malicious vehicles can launch the Sybil attack to impersonate the legitimate sensing vehicles to obtain benefits. In this regard, GAN-based defense scheme \textcolor[rgb]{0.00,0.00,0.00}{was introduced} to recognize the Sybil attack, where GAN \textcolor[rgb]{0.00,0.00,0.00}{was} used to detect anomaly trajectory among vehicles. \textcolor{black}{In the eight different fake trajectories classification experiments, the Area Under the Receiver Operating Characteristic curve (AUC-ROC) score can always range from 90\% to 97\%. Hence, the effectiveness of the GAN-based approach is evaluated.}

\textbf{Generative diffusion model (GDM).} GDM can generate data that is similar but not identical to the training data by learning input training \textcolor[rgb]{0.00,0.00,0.00}{samples}. Generally, inspired by diffusion process of nonequilibrium thermodynamics, GDM can define the diffusion steps, which aim to gradually add random noise to the data, and then learn to reverse the \textcolor[rgb]{0.00,0.00,0.00}{process through denoising}. Finally, the \textcolor[rgb]{0.00,0.00,0.00}{required} data samples can be constructed from the noise. Besides, GDM evolves during the fixed process, where the latent variables are with the same high dimensionality as the original data. In the data collection process of SPPMCS, GDM can be used to detect abnormal data to realize the network \textcolor[rgb]{0.00,0.00,0.00}{intrusion} detection. For example, in \cite{tang2023diffusion}, to improve the low classification performance of intrusion detection due to data imbalance, GDM \textcolor[rgb]{0.00,0.00,0.00}{was} adopted to generate the rare class of intrusion attacking data.
\textcolor{black}{For rare class abnormal data, simulation results show that the GDM-based rare data balancing scheme can enhance the classification accuracy from 30\% to 65\%, and also perform better than non-over-sampling and over-sampling baselines.}

\textbf{Transformer-based model (TBM).} TBM is \textcolor[rgb]{0.00,0.00,0.00}{widely} applied in natural language processing tasks such as translation, summarization, and text generation. \textcolor[rgb]{0.00,0.00,0.00}{TBM was designed to be more efficient than traditional recurrent neural networks in processing long sequences of data.} Specifically, in traditional recurrent neural networks, each input needs to be processed sequentially, which \textcolor[rgb]{0.00,0.00,0.00}{leads} to loss of information. In contrast, TBM processes sequential data by introducing an attention mechanism that dynamically distributes attention between different positions in the sequence, greatly \textcolor[rgb]{0.00,0.00,0.00}{improving} the data process efficiency. TBM can be used to detect malicious source codes of the sensing data-driven software in SPPMCS. For example, in \cite{tsfaty2022malicious}, based on the Transformer-based deep learning method, the real-world code injection cases to source code packages can be detected. \textcolor{black}{Specifically, after training a transformer-based model, named Code2Seq, with the function dataset, anomaly detection techniques can be applied on Code2Seq's representation for each function type within the code. Extensive experiments have demonstrated that TBM can effectively identify functions that are injected with malicious code with 90.9\% detection accuracy.}

We summarize and compare these generative AI models, including architecture, description, components, advantages, disadvantages, and applications in SPPMC in Table \ref{Summarization}.

\section{Generative AI-enabled SPPMCS}
\begin{figure*}[!htb]
  \centering
  \includegraphics[width=17cm]{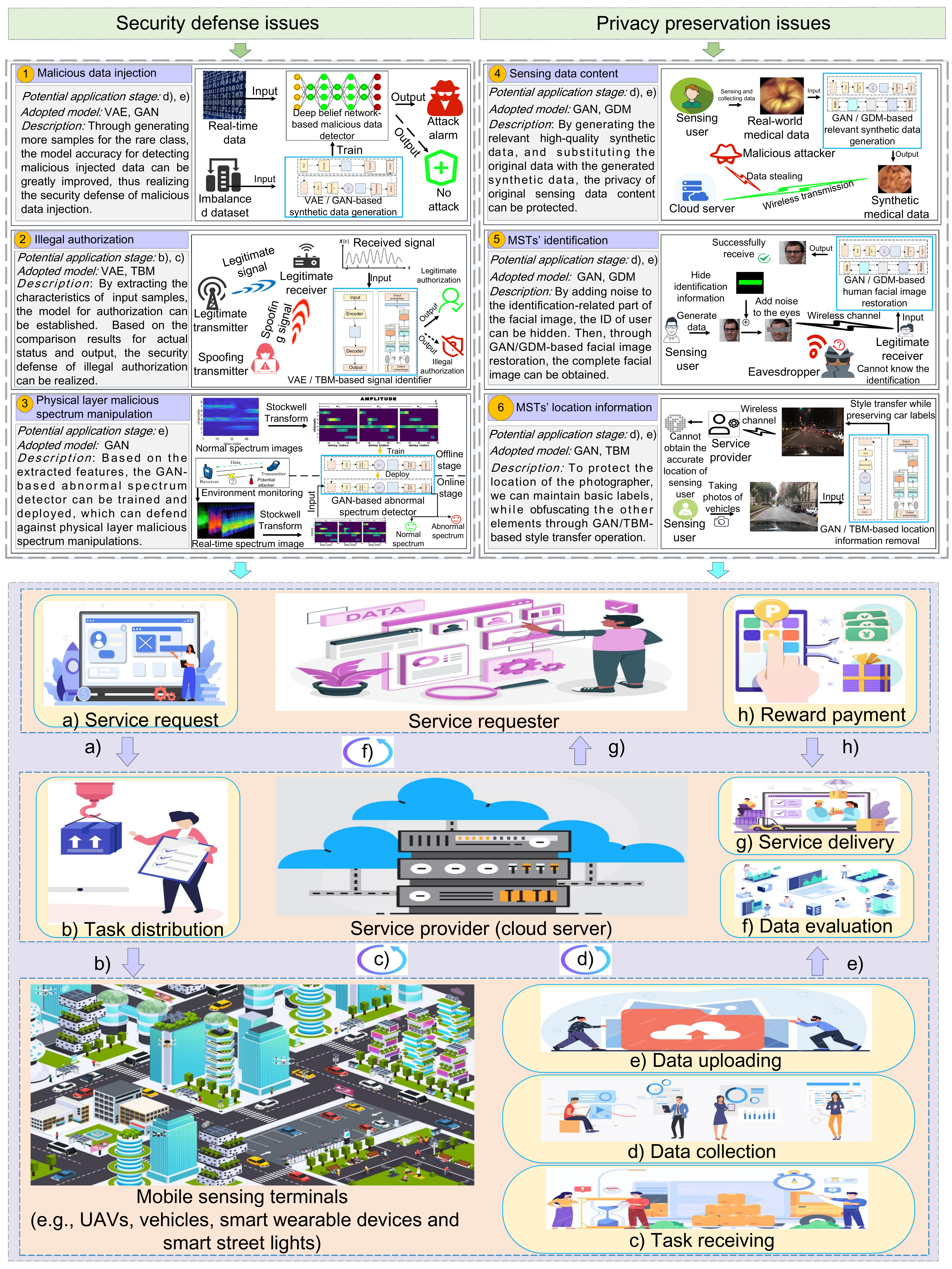}\\
  \caption{\textcolor{black}{The role of generative AI in security defense and privacy preservation aspects for SPPMCS applications. In the security defense aspect,   generative AI can help handle malicious data injection, illegal authorization, and physical layer malicious spectrum manipulation on the task distribution, task receiving, data collection, and data uploading stages. As for the privacy preservation issue, on the data collection and data uploading stages, the generative AI-enabled SPPMCS can alleviate the privacy preservation concerns on sensing data, and MST's identification and location information.}}
  \label{structure}
\end{figure*}


\subsection{Research focuses and solutions}

As shown in Fig. \ref{structure}, generative AI-enabled SPPMCS can greatly improve the system security performance, and enhance the privacy preservation effect. \textcolor[rgb]{0.00,0.00,0.00}{The} major research focuses and solutions of generative AI-enabled SPPMCS \textcolor[rgb]{0.00,0.00,0.00}{are} summarized as follows.

1) \textbf{Security Defense of Malicious Data Injection.} Traditional discriminative AI, like CNN or long short-term memory (LSTM), detects sensing data flow for identifying malicious injection behavior \cite{he2017real}. However, malicious data is injected in a discontinuous, sudden, and irregular manner, which can make abnormal data belong to the rare class samples. In this regard, when training the traditional discriminative AI-based models, data imbalance issues can affect the model accuracy. In contrast, generative AI models, such as VAE and GAN, address this by generating synthetic samples of rare class for training \cite{liu2022intrusion}. For instance, using VAE for data augmentation in \cite{liu2022intrusion} achieved a 98.557\% accuracy in detecting malicious attack data after balancing rare class samples. GAN, similarly, improves the detection model's performance by creating synthetic abnormal data, achieving a balanced training dataset.

2) \textbf{Security Defense of Illegal Authorization.} Traditional non-AI methods, like encryption and digital signatures, are used to combat identity forgery and illegal authorization in SPPMCS. However, these approaches may be hindered by key management costs and slow adaptability in dynamic environments (e.g., MSTs joining or existing in the current sensing task). Generative AI, specifically VAE and TBM, offers physical-layer authentications and dynamic trust evaluation solutions. For instance, by treating the channel characteristic of MSTs as the authentication mark in the small samples training process, after extracting the channel impulse response (CIR) characteristics and improving the representation ability of the CIR, VAE can output the MST¡¯s authentication results without knowing malicious attacker¡¯s channel information in advance \cite{meng2022physical}. Additionally, TBM constructs a trust model by analyzing participants' time characteristics, predicting future behavior, and evaluating trustworthiness based on the similarity between predicted and actual behavior \cite{yu2022dynamic}.

\begin{table*}[t]
\centering
\caption{\textcolor{black}{Comparison among non-AI, traditional discriminative AI, and generative AI technologies in SPPMCS.}}  \label{Summarization2}
\resizebox{0.92\width}{!}{
\begin{tabular}{c}
\includegraphics[width=19cm]{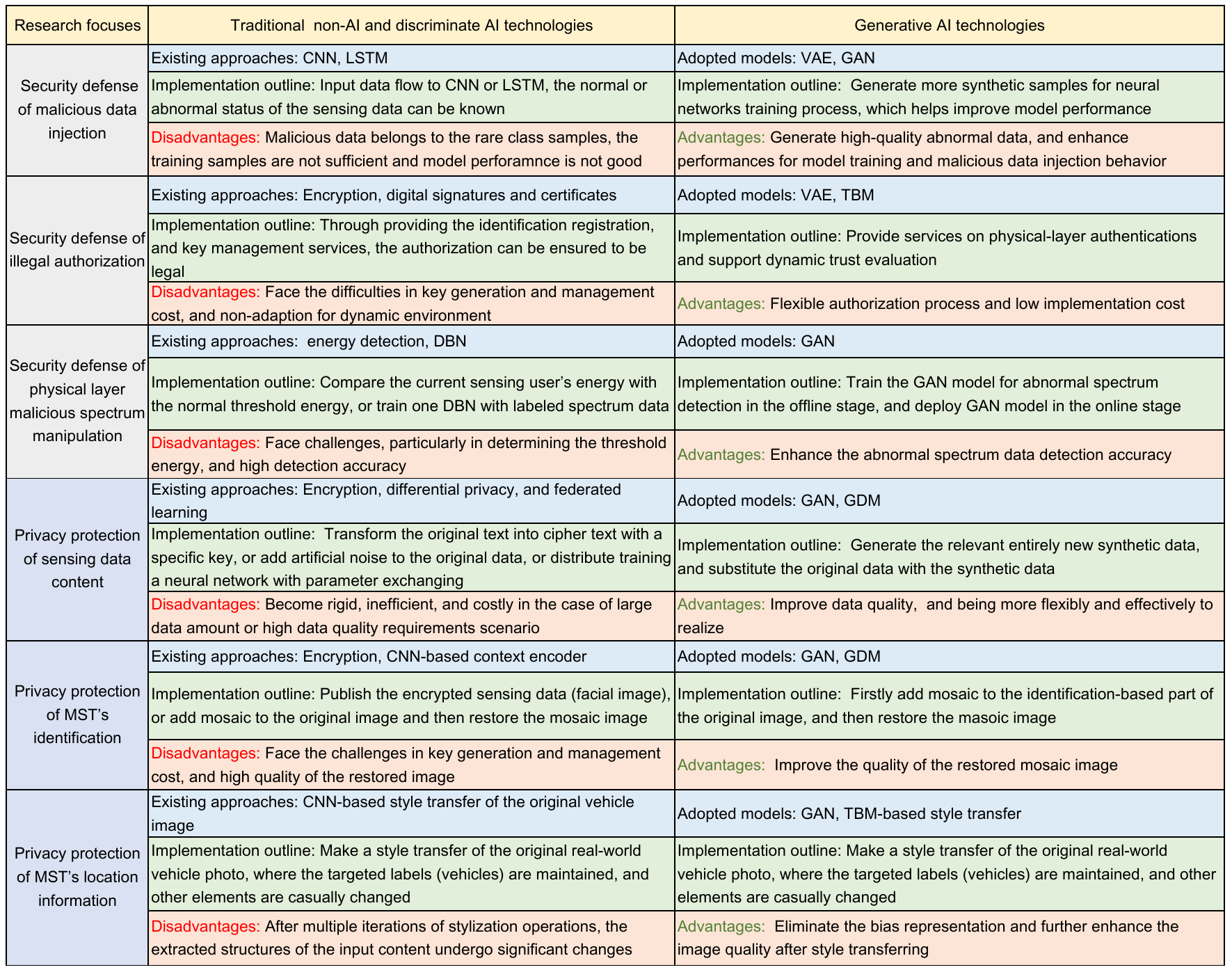}\\
\end{tabular}}
\end{table*}

3) \textbf{Security Defense Against Malicious Spectrum Manipulation at the Physical Layer.} With limited spectrum resources, potential attackers manipulate radio spectrum to misguide sensing users and compromise data freshness. Traditional non-AI and discriminative AI methods, like energy detection technique and deep belief network (DBN)-based malicious spectrum detection model, face challenges in establishing thresholds and ensuring high accuracy. In contrast, GAN offers a unique approach to recognize abnormal spectrum data and defending against physical layer spectrum manipulation. In offline training, the GAN model is trained on abnormal spectrum features extracted from Stockwell transform (ST)-based images. During online testing, the abnormal spectrum can be detected by applying an ST operation to real-time spectrum images and inputting the resulting amplitude-time image into the deployed GAN. Notably, testing on 1200 image samples revealed that the GAN-based detection approach achieved a detection accuracy of 97.46\%, surpassing the DBN-based detection method (96.77\%) and the energy detection technique (96.67\%) \cite{8976292}.

4) \textbf{Privacy Protection of Sensing Data Content.} Traditional methods, like encryption and differential privacy, protect sensing data content by transforming original text into ciphertext using a specific key or introducing artificial noise to the initial data. Additionally, conventional discriminative AI approaches, like federated learning, facilitate the exchange of neural network parameters between a local model and a central server, reducing the need for direct data interaction. However, these methods face challenges in data reversibility, adaptability to complex data, and flexibility concerning varying privacy requirements.
In contrast, training GAN and GDM models with the
original data, synthetic data can be generated to replace the real data. For instance, for tasks like CNN-based abnormality detection in medical imaging, due to the patient's data protection legislation, the original medical images cannot be directly used for training the CNN model. In this regard, GAN help generate fake but realistic medical images for training the CNN-based detector model. Experiments showed the CNN model trained on synthetic data achieved 79.1\% AUC-ROC, while real-world data achieved 90.9\% \cite{8941751}.

5) \textbf{Privacy Protection of MST's Identification.} In the MCS recruitment stage, MSTs upload facial photos for legal identification \cite{yangppfo} at the SP\footnote{https://dzone.com/articles/modern-web-applications-authentication-using-face}. The original facial image, containing MST identification, risks leakage from malicious eavesdroppers employing inferring attacks. Traditional encryption-based methods face challenges in key generation, management costs, and inflexible adaptation to different identification privacy preservation requirements.
Discriminative AI, like CNN-based context encoders, initially adds mosaic (noise) to the identification-related part of the facial image (e.g., eyes). Then, the SR restores the image using a CNN-based context encoder to complete MST authentication. However, the CNN-based context encoder technique may introduce undesirable artifacts and lacks high-frequency details \cite{javed2019umgan}, potentially affecting
MST authentication results.
Generative AI, specifically GAN and GDM, effectively addresses these concerns through high-quality image restoration\footnote{Compared with the privacy protection of sensing data content, since only facial photos are permitted, and synthetic facial images are not allowed in the registration and authentication process, identification information protection cannot be realized by substituting the original facial photo with a synthetic one.}. Testing with CelebA Face dataset shows GAN-based approaches outperform than CNN-based one, with improved 481 Mean Squared Error (MSE), 0.853 Structural Similarity Index (SSIM), and 26.68dB Peak Signal-to-
Noise-Ratio (PSNR) indicators \cite{javed2019umgan}.

6) \textbf{Privacy Protection of MST's Location.} When an SR (e.g., traffic police) needs to verify if vehicles on the road have license plates, MSTs (e.g., vehicles) can execute sensing tasks like taking photos. However, real-world photos pose privacy risks\footnote{https://www.npr.org/transcripts/1219984002}, potentially leaking users' location information. To address this, a promising idea is style transfer for the original vehicle photo, maintaining targeted labels (e.g., vehicles) while casually changing other elements (e.g., streetlights). The style-transferred photo is entirely new, preventing location analysis. Traditional CNN-based models can be used for style transfer, but this method has a biased representation issue, where after multiple  iterations of stylization operations, the extracted structures of the input content undergo significant changes. In contrast, generative AI, especially GAN and TBM, eliminates bias and enhances image quality after style transfer\footnote{Compared with sensing data content privacy, synthetic vehicle images don't contribute to license plate checking, so location protection requires more than substituting the original with synthetic photos.}. Numerical results show TBM-based style transfer achieving a content perceptual loss of 1.91, while other typical CNN-based models, including AdaIN, Avatar, SANet, AAMS, and MAST, exhibit content perceptual losses of 2.34, 2.84, 2.44, 2.44, and 2.46, respectively \cite{deng2022stytr2}.

In summary, in the security enhancement and privacy preservation improvement aspects of SPPMCS, the comprehensive comparison among non-AI, traditional discriminative AI, and generative AI technologies is presented in Table II.
\textcolor{black}{Note that even though generative AI can effectively help defend against security attacks and address privacy concerns, it is also with the following limitations. 1) The implementation of GAI-based privacy-preserving schemes may require high computing resources and consume much energy, increasing the training cost. 2) The synthetic data-based privacy-preserving techniques may face difficulties in being regulated by relevant laws and ethics. 3) Due to a large amount of training data and long training time, the GAI-based privacy-preserving methods may encounter a response delay.}

\subsection{Potential challenges}

As generative AI brings lots of benefits for SPPMCS, the security defense against malicious attacks can be realized, and the comprehensive privacy issues can also be addressed. However, as SPPMCS contains lots of participants with various requirements, and involves several different data interaction processes, the implementation cost and timeliness guarantee could be the potential challenges. To be specific, due to the high quality requirement of new synthetic data generation and sensing data analysis, the implementation cost of integrating generative AI into SPPMCS could increase the computation resource consumption and response time of the attack defending scheme.

\section{Sensing data content protection in generative AI-enabled SPPMCS}

\begin{figure*}[!htb]
  \centering
  \includegraphics[width=18cm]{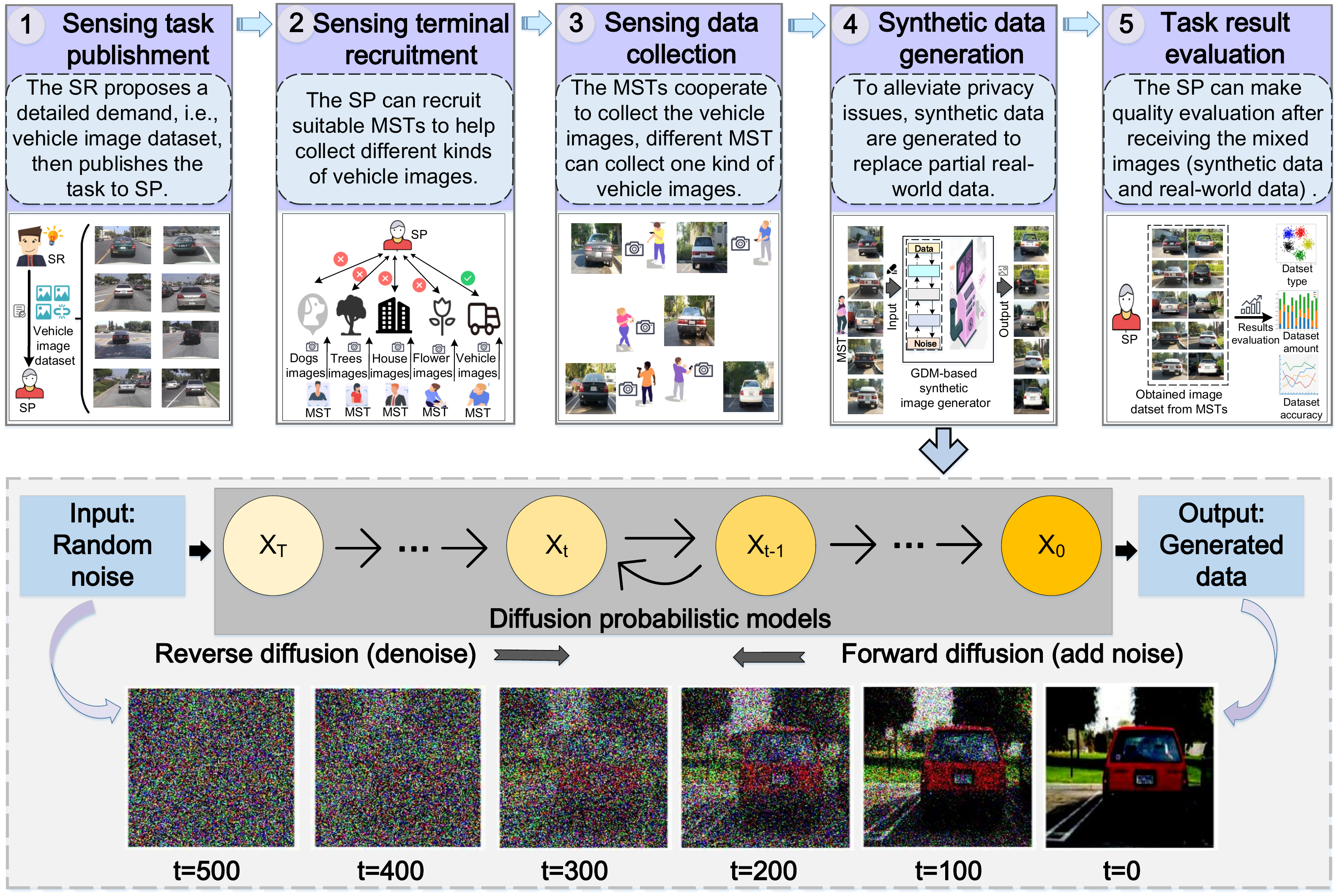}\\
  \caption{\textcolor{black}{A framework for sensing data content protection in generative AI-enabled SPPMCS, where the GDM model is adopted to generate synthetic vehicle images. In the GDM-based synthetic image generation process, the feature of an original image can be learned by adding noise in the forward diffusion stage. On this basis, denoising operation can generate random noise to the target image in the reverse diffusion stage. Since synthetic data contains no privacy information of the real-world, and by substituting real-world images with synthetic images, the sensing data privacy issues can be alleviated.}}
  \label{framework}
\end{figure*}


\subsection{Research motivation}

Under the reward payment-based incentive mechanism on recruitment, MSTs are willing to collect required sensing data to help the SR to establish one dataset. However, the collected sensing data contains private information, and some AI technologies can be maliciously used to leak privacy. In this regard, directly uploading the original photo can lead to privacy breaches and even violations of the law\footnote{https://themarkup.org/the-breakdown/2020/03/12/photos-privacy}. For example, the publication of vehicle photos on the road may have privacy implications, as they may contain sensitive information such as the exact location of the vehicle and its license plate number.
Therefore, careful consideration needs to be given to privacy protection when publishing such photographs, and it is necessary to take measures to prevent privacy breaches.
Moreover, such a privacy leakage risk can also seriously influence MSTs' motivation for participating in a sensing task.
\textcolor{black}{In this regard, to achieve effective dataset augmentation, dataset balance, and privacy preservation, we employ generative AI to generate high-quality synthetic data.
By substituting collected real-world sensing data with relevant generated data, the privacy issues of the original sensing data can be alleviated\footnote{https://www.statice.ai/post/how-manage-reidentification-risks-personal-data-synthetic-data}, and the sensing \textcolor{black}{data quality\footnote{\textcolor{black}{Note that the quality of data generated by generative AI depends on original data's quality, adopted generative AI models, and training environment three aspects, simultaneously.}} can be ensured simultaneously.}}

\subsection{Proposed framework}

With generative AI, the sensing data content privacy preservation issues can be addressed in SPPMCS. Accordingly, as shown in Fig. \ref{framework}, based on the GDM model\footnote{https://arxiv.org/abs/2006.11239}, we propose the framework for the sensing data content preservation in generative AI-enabled SPPMCS, which mainly consists of the following five key steps.

\textbf{Step 1: Sensing Task Publishment.} As an initiator of a sensing task, an SR first has one specific application demand, which can be realized based on corresponding sensing data. Then, owing to the limited data processing capacity for collection, analysis, and utilization, the SR outsources the application demand to the SP, i.e., publishing the detailed sensing data collection requirements to the SP.

\textbf{Step 2: Sensing Terminal Recruitment.} After the SP receives the detailed requirement from an SR, it can recruit suitable MSTs to participate in the sensing data collection task. In addition, the SP should also clearly declare the reward payment rule to the MSTs. The more high-quality data and the better the SR's demand effect can be realized, the more reward would be paid. Under such a designed incentive mechanism, the willingness and fairness of MSTs can be ensured.

\begin{figure*}[t]
  \centering
  \includegraphics[width=18cm]{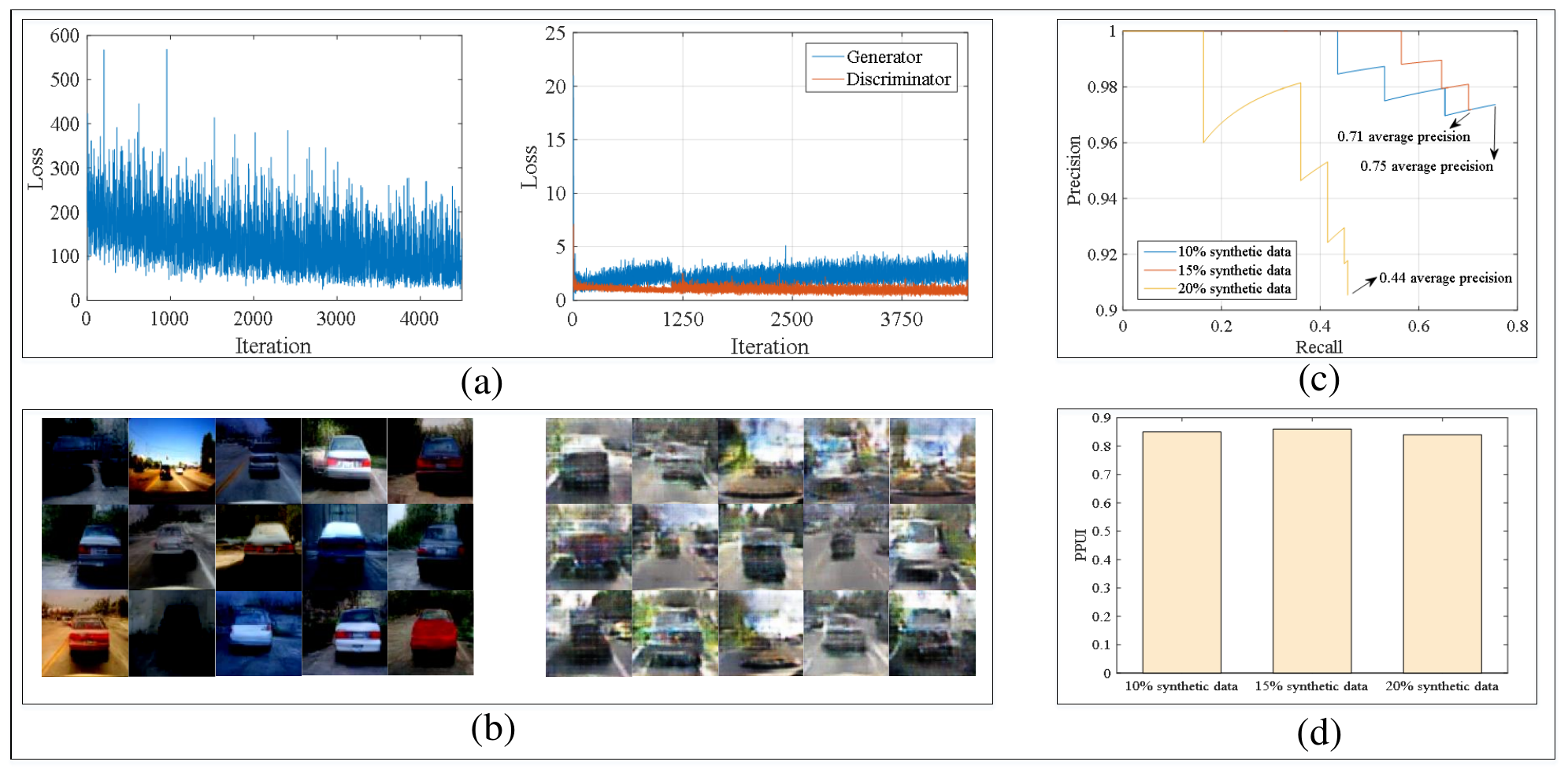}\\
  \caption{Performance evaluation of the proposed framework. (a) Training process of GDM and baseline models. Left: GDM. Right: GAN. (b) Images generated by GDM and baseline models. Left: GDM. Right: GAN.  (c) Precision recall performance for YOLOv3 detection model with different percentage of synthetic data for training dataset. (d) Privacy-preserving utility index performance with different percentage of synthetic data for training dataset.}
  \label{simulation2}
\end{figure*}

\textbf{Step 3: Sensing Data Collection.} According to the requirement from SR, the MSTs can collect the needed sensing data. Note that due to the different sensing data collection capacities, different MSTs can choose to finish the sensing task individually or cooperatively. For example, when an SR needs five types of sensing data, some MSTs can collect all required kinds of sensing data, they can finish the sensing task individually. Meanwhile, other MSTs can only collect partial types of sensing data, so they need to cooperate to collect all required sensing data. Although cooperation can improve the completeness of sensing task execution, it can also lead to the reward being divided up.

\textbf{Step 4: Synthetic Data Generation.} MSTs may not submit all the collected real-world sensing data for privacy preservation consideration. With the help of GDM, relevant high-quality synthetic data, which contains no private information, can be generated. Therefore, to simultaneously ensure the submitted sensing data quality and protect privacy, the MSTs can upload mixed sensing data, including synthetic and partial real-world collected data.

\textbf{Step 5: Task Result Evaluation.} The SP can evaluate data quality after receiving the mixed sensing data from the MSTs, e.g., type, amount, and accuracy. Then, each MST can be paid the corresponding reward based on the predefined incentive mechanism, i.e., the effectiveness verification results of mixed data quality.

\subsection{Case study}

In this case study, we take an example of vehicle image collection task,
where the SR needs to establish a vehicle image dataset. Such a required dataset can be used to train the vehicle detection model for the autonomous driving scenario. 
Specifically, the SP can recruit the MSTs to collect the vehicle images, and MSTs submit their real-world collected sensing data and partial synthetic data to finish the sensing task.
Generally, in the numerical results, we evaluate the feasibility and effectiveness of the proposed framework on four aspects: 1) the training process of the GDM and baseline (GAN) models, 2) the visual effect of generated synthetic vehicle images, \textcolor{black}{3) how much synthetic data can be mixed to vehicle detection model's requirements, 4) how to quantitatively measure the effectiveness of privacy preservation and usefulness of the synthetic data.}
By analyzing the numerical results, we aim to provide significant insights for generative AI-enabled SPPMCS  corresponding to the above focused aspects.

The simulation experiments are based on the GAN\footnote{https://ww2.mathworks.cn/help/deeplearning/ug/train-generative-adversarial-network.html}, GDM\footnote{https://ww2.mathworks.cn/help/deeplearning/ug/generate-images-using-diffusion.html}, and YOLOv3\footnote{https://ww2.mathworks.cn/help/deeplearning/ug/object-detection-using-yolo-v3-deep-learning.html} models, 
which are implemented on Matlab R2023a. Besides, this case study also employs a compact labeled dataset comprising 295 images. A significant portion of these images is sourced from the Caltech Cars 1999 and 2001 datasets, accessible on the Caltech Computational Vision website, originally curated by Pietro Perona and employed here with permission. Each image within the dataset features one or two labeled instances of a vehicle. The training and evaluation operations for the above models are executed on a desktop PC with Intel(R) Core(TM) i7-11700 CPU @ 1.60GHz, NVIDIA GeForce RTX 2060 GPU, 16.0 GB memory and Windows 10 home operation system.

In Fig. 3(a), we present the training processes for the GDM and the baseline (GAN) models. The parameter settings include a mini-batch size of 32 for 500 epochs, a learning rate of 0.005, and a squared gradient decay factor of 0.9999. As shown in Fig. 3(a), both GDM and GAN converge to steady loss points with increasing training steps, affirming the feasibility of the proposed framework¡ªdemonstrating generative AI's ability to generate synthetic data for privacy preservation. Notably, GAN exhibits a faster training speed compared to GDM; for instance, training the GDM and GAN models in this experiment took 12 hours 38 minutes, and 23 minutes, respectively.

Fig. 3(b) shows 15 synthetic vehicle images (64$\times$64 pixels). GDM outperforms the GAN model in generating higher-quality and diverse images due to its incremental learning approach. Unlike GAN, GDM progressively learns image characteristics through step-wise noise introduction or denoising, capturing more details for superior effects. GDM effectively generates diverse synthetic data with enhanced creativity, while GAN excels in generating data similar to the existing dataset. From a privacy protection perspective, GDM's incremental learning enhances the complexity and authenticity of generated images, ensuring the utmost similarity to the originals. Substituting original images with synthetic ones effectively mitigates potential privacy risks.

In Fig. 3(c), YOLOv3's performance is shown with varying proportions of GDM-generated synthetic data in the training dataset. An increase in the proportion of GDM-generated synthetic data correlates with a decrease in YOLOv3's detection precision. While synthetic data safeguards original data from privacy leakage, its integration can impact downstream tasks, influencing model accuracy. Defining an optimal percentage of synthetic data for simultaneous privacy protection and downstream task accuracy maintenance is crucial, requiring the introduction of a relevant new metric.

In Fig. 3(d), \textcolor{black}{a novel metric, the Privacy-Preserving Utility Index (PPUI), is introduced to assess the effectiveness of privacy preservation and synthetic data utility. PPUI is calculated by adding the synthetic data percentage of the training dataset to the average accuracy of the downstream task model, both ranging from 0 to 1. PPUI is influenced by the synthetic data percentage, where higher percentages enhance privacy preservation but may lead to decreased model accuracy, resulting in a lower PPUI. The strategy with the highest PPUI represents the optimal balance between privacy preservation and downstream task completion.} For instance, as illustrated in Fig. 3(d), incorporating 15\% synthetic data results in a 0.86 PPUI, striking a balance between privacy protection and model accuracy that proves effective for subsequent tasks.

\section{Open issues}

\subsection{Model interpretability}

Since generative AI is directly employed for security defense and privacy protection in SPPMCS, it deals with real-world data that contains a wealth of sensitive and crucial information about individuals. Therefore, enhancing the interpretability of generative AI to render the model transparent and understandable can instill greater trust and support. Furthermore, the interpretability of generative AI models can assist in identifying concealed issues or bugs, contributing to the continuous improvement of the performance of generative AI-enabled SPPMCS\footnote{https://techxplore.com/news/2023-12-prompts-chatgpt-leak-private.html}.

\subsection{Lightweight model optimization}

With the growing demand for implementation timeliness and cost-effectiveness in generative AI-enabled SPPMCS, it is crucial to optimize the adopted model for lightweight \textcolor{black}{applications for resource-limited devices}. This optimization can play a significant role in reducing computation resources, improving algorithm response speed, and enhancing data processing efficiency. Additionally, operations such as model compression, pruning, quantification, and parallel computing, can further contribute to a substantial reduction in model complexity.

\subsection{\textcolor{black}{Privacy regulation and ethical assurance}}

\textcolor{black}{Integrating generative AI into SPPMCS can cause certain concerns from the privacy regulation and ethical aspects. In this regard, several measures should be taken. For example, from the privacy regulation aspect, we can provide explainability about the synthetic data generation process, enabling data protection authorities to verify compliance with privacy regulations. From the ethical assurance aspect, to assess the potential ethical impact, social impact assessments need to be conducted prior to the deployment of synthetic data-driven applications.}

\section{Conclusion}
In this paper, we delved into the integration of generative AI and SPPMCS to further address security and privacy issues. This integration involves generating new synthetic data to replace the original data for analysis and processing. We introduced the preliminaries of SPPMCS. Subsequently, we presented typical generative AI models, VAE, GAN, GDM, and TBM. The discussion then explores potential research focuses, solutions, and challenges of generative AI-enabled SPPMCS. Additionally, we proposed a framework for sensing data content protection in the generative AI-enabled SPPMCS, with a case study evaluating its performance. Finally, the paper concludes by highlighting open issues related to generative AI-enabled SPPMCS.

\bibliography{ref}{}
\bibliographystyle{IEEEtran}

\end{document}